\documentclass[review]{elsarticle}
\usepackage{graphicx} 

\usepackage{graphicx}
\begin{document}

\begin{frontmatter}

\title{Non-equilibrium diffusion characteristics of a particle system and the correspondence to a social system}

\author[mymainaddress]{Peng Wang}
\author[mymainaddress]{Feng-Chun Pan}
\author[mymainaddress,mysecondaryaddress]{Jie Huo}
\author[mymainaddress,mysecondaryaddress]{Xu-Ming Wang\corref{mycorrespondingauthor}}
\cortext[mycorrespondingauthor]{Corresponding author}
\ead{wang$\_$xm@126.com}

\address[mymainaddress]{School of Physics and Electronic-Electrical Engineering, Ningxia University, Yinchuan 750021, PR China}
\address[mysecondaryaddress]{Ningxia Key Laboratory of Intelligent Sensing for Desert Information, Yinchuan 750021, PR China}

\begin{abstract}
A Langevin equation is suggested to describe a system driven by correlated Gaussian white noise as well as with positive and negative damping demarcated by a critical velocity. The equation can be transformed into the Fokker-Planck equation by the Kramers-Moyal expansion. The solution exhibits some non-equilibrium phenomena. In the beginning the distribution curve of velocity/energy takes on a random oscillation, and then a near-equilibrium distribution described by the Boltzmann distribution is gradually established. However, a spike appears on the distribution curve and breaks this stable distribution. The spike moves in the direction of velocity/energy decreasing and is nonlinearly enlarged so as to sustain. The final distribution is a sharp peak formed by a monotonically ascending segment and a monotonically descending one. The calculating results of the statistical quantities demonstrate that the process is a sub-diffusion, and the spike originates from the correlation between noise and space. Based on a basic hypothesis that the generalized displacement is regarded as the information carried by an opinion particle, and the velocity is taken as the sensitivity of an agent to an event, we map the observations in this system to a social system to understand the propagation of public opinion or message. 
\end{abstract}


\end{frontmatter}


\section{\label{sec:level1}Introduction}

Random diffusion is a basic subject for statistical physics and relevant fields, such as the flow in porous media, particle diffusion in turbulent field and so on\cite{megen1986,megan1987,risken1984,notarnicola2018,kula1998,astumian1994,carcaterra2016,Salgado2019}. It is well known that the diffusion is caused by the gradient of macroscopic quantity, density, while it is actualized by the collision among particles at a micro level. The mean-field methods, such as lattice Boltzmann models\cite{sun2016,sun2013400,safari2013,bhatnagar1954,gan2015}, continuous-time random walks model\cite{diniz2017,silva2013,hara1979,schutz2004} and generalized Langevin model\cite{das2006,friedrich2006}, have made some important contributions to investigating the phenomena. The calculating results in Refs.\cite{Salgado2019,diniz2017,das2006} indicate that the diffusion coefficient described by the mean square displacement is of power scaling. The value of the scaling exponent equal to, greater than and less than $1$ represents free diffusion, super-diffusion and sub-diffusion, respectively. As everyone knows that random fluctuation plays an important role in the non-equilibrium diffusion. The Langevin equation\cite{notarnicola2018} is the simplest and main method to discuss the effects of internal or external fluctuation. It is generally believed that random noise is unfavorable to an orderly pattern. However, some times the opposite is ture---the random noise will cause some novel phenomena, for instance, non-equilibrium transition and transport in a nonlinear system\cite{astumian1994}. Therefore, a comprehensive study on the effect of noise in a non-equilibrium system will be conducted to reveal the mechanism.

From the perspective of statistical physics, the social system can be regarded as the one of quasi-particle system, for instance, opinion particle. So the social phenomena such as the propagation of public opinion\cite{Gaudiano2019,Pinto2019}, the diffusion of message, etc., can be described by the transport of such qusi-particles\cite{Zhu2017,Pineda2009}. Physicists have made many attempts and efforts. State change of an opinion particle is investigated in the framework of the Newtonian Mechanics by using probabilistic descriptions of the variables based on Bayesian theorem\cite{martins2014}. Collective opinion formation is modelled by active Brownian particles via introducing master equation\cite{Schweitzer2000}. A transition between order and disorder of opinion dynamics induced by external noise is found in the Deffuant model approximated by master equation\cite{Pineda2009,Khalil2019}. The effect of repulsive interaction between the opinion particle in the Deffuant model is simulated in a similar way\cite{martins2010}.

This paper investigates the random diffusion in a non-equilibrium system with positive and negative damping. We aim to reveal its characteristics and the corresponding mechanism, and then map them to a social system to understand some of the propagation behaviors. It is interesting that a non-equilibrium spike will appear on the near-equilibrium distribution curves of velocity and energy. The calculating results reveal that the spike is induced by the correlation between the noise and the space. The spike moves in the direction of velocity/energy decreasing due to the dissipation while it is enlarged nonlinearly. These findings will be mapped to a social system to describe some of its behaviors. The pumping(negative damping) mechanism could be regarded as the external influence or guiding learning help provided for the individuals, of poor-learning ability or insensitive to things, to efficiently improve their ability. The dissipative(positive damping) mechanism could be taken as that the social "noise" such as some comments, concerns, lies and slanders, etc. can be nonlinearly enlarged, although the system are dissipated in whole. 

\section{\label{sec:level2}System and its description by Fokker-Planck equation}

The random diffusion of particles can be described by the following equations,
\begin{equation}\label{eq:eq1}
\left\{\begin{array}{lll}
&\dot x = v  \\
&\rho\dot v = - \gamma (v)v - \nabla U(x) + G(v)\xi (t)\\
\end{array}\right.
\end{equation}
where $\rho$ is the particle density, $v$ the velocity, $x$ the displacement and $U(x)$ the generalized potential. To simplify the model, we consider that the gradient of generalized potential is linear. So the generalized potential can be written as
\begin{equation}\label{eq:eq2}
U(x) = \frac{\omega _0^2}{2}{x^2}
\end{equation}
where $\omega _0$ is constant.

In Eq.\ref{eq:eq1}, $\gamma(v)$ is the damping function of which the value might be positive or negative in some systems. The positive damping describes a passive motion of particles, while the negative damping means the active motion of particles. In a real system, energy replenishing is needed to maintain the motion of particles on one hand and energy dissipating is natural and inevitable on the other hand. In fact, these inferences are supported by the investigations executed by Rayleigh and Helhmhplta based on the so-called nonlinear friction model. They pointed out that there is both positive and negative damping in a micro system with very slow particle\cite{strutt2009,helmholtz1954}. Here we adopt their suggestion, that is, 
\begin{equation}\label{eq:eq3}
\gamma(v)={\gamma _0}({v^2}-\alpha),
\end{equation}
where $\alpha$ is a determined positive constant. The negative $\gamma(v)$ when $v$ is less than $\sqrt\alpha$ means active motion, which denotes the pumping diffusion. On contrary, the positive $\gamma(v)$ as $v$ is greater than $\sqrt\alpha$ denotes the dissipation diffusion.

The third term in Eq.\ref{eq:eq1} is noise, where $\xi(t)$ denotes the Gaussian noise, $G(v)$ its strength. It is composed of additive noise $\zeta(t)$ and multiplicative noise $g(v)\varsigma (t)$($g(v)$ is the strength of multiplicative noise). Therefore, Eq.\ref{eq:eq1} becomes
\begin{equation}\label{eq:eq4}
\left\{\begin{array}{lll}
\dot x = v      \\
\dot v=-\gamma (v)v -\nabla U(x) + g(v)\varsigma (t) + \zeta (t)    \\
\end{array}\right.
\end{equation}
where $\varsigma (t)$ and $\zeta (t)$ are the Gaussian white noise that can be characterized as
\begin{equation}\label{eq:eq5}
\left\{\begin{array}{lll}
\left\langle {\varsigma (t)} \right\rangle  = 0\\
\left\langle {\zeta (t)} \right\rangle  = 0\\
\end{array}\right.
\end{equation}
and
\begin{equation}\label{eq:eq6}
\left\{\begin{array}{lll}
\left\langle {\varsigma (t)\varsigma (t')} \right\rangle  = 2D\delta (t - t')\\
\left\langle {\zeta (t)\zeta (t')} \right\rangle  = 2Q\delta (t - t')\\
\left\langle {\varsigma (t)\zeta (t')} \right\rangle  = \left\langle {\zeta (t)\varsigma (t')} \right\rangle  = 2\theta \sqrt {DQ} \delta (t - t')
\end{array}\right.
\end{equation}
where $D$ and $Q$ denote the autocorrelation strength coefficient of additive noise and that of multiplicative noise, respectively. $\theta$ is the correlation strength between the two kind of noises and satisfies with $0 \le \theta  \le 1$. The second order noise correlation can be written as
\begin{equation}\label{eq:eq7}
\begin{array}{lll}
{\left[ {G(v)} \right]^2}\left\langle {\xi (t)\xi (t')} \right\rangle &= \left\langle {\left( {g(v)\varsigma (t) + \zeta (t)} \right)\left( {g(v)\varsigma (t') + \zeta (t')} \right)} \right\rangle \\
 &= \left\langle {{{\left[ {g(v)} \right]}^2}\varsigma (t)\varsigma (t') + g(v)\varsigma (t)\zeta (t') + g(v)\varsigma (t')\zeta (t) + \zeta (t)\zeta (t')} \right\rangle \\
 &= {\left[ {g(v)} \right]^2}\left\langle {\varsigma (t)\varsigma (t')} \right\rangle  + g(v)\left\langle {\varsigma (t)\zeta (t')} \right\rangle  + g(v)\left\langle {\varsigma (t')\zeta (t)} \right\rangle  + \left\langle {\zeta (t)\zeta (t')} \right\rangle \\
 &= \left( {2D{{\left[ {g(v)} \right]}^2} + 4\theta \sqrt {DQ} g(v) + 2Q} \right)\delta (t - t')
 \end{array}
\end{equation}
So the correlation strength of noise is
\begin{equation}\label{eq:eq8}
G(v) = {\left( {2D{{\left[ {g(v)} \right]}^2} + 4\theta \sqrt {DQ} g(v) + 2Q} \right)^{\frac{1}{2}}}
\end{equation}
Consequently, the Langevin equation, Eq.\ref{eq:eq4}, can be transformed into the Fokker-Planck equation
\begin{equation}\label{eq:eq9}
\begin{array}{lll}
\frac{{\partial P(x,v,t)}}{{\partial t}} =& \frac{\partial }{{\partial v}}\left( {\frac{1}{\rho }\left( {\gamma (v)v + \nabla U(x)} \right) - \frac{1}{{{\rho ^2}}}G(v)G'(v)} \right)P(x,v,t)\\
&+ \frac{1}{{{\rho ^2}}}\frac{\partial }{{\partial v}}\left( {{{\left[ {G(v)} \right]}^2}\frac{{\partial P(x,v,t)}}{{\partial v}}} \right) - \frac{\partial }{{\partial x}}\left( {vP(x,v,t)} \right)
\end{array}
\end{equation}
where $P(x,v,t)$ is the probability to find a particle with $v$ at time $t$ and $x$. Please refer to the detailed derivation process of the transformation from Eq.4 to Eq.9 in Ref.\cite{risken1984}.

According to Stratonovich hypothesis\cite{risken1984}, Eq.\ref{eq:eq9} can be transformed into
\begin{equation}\label{eq:eq10}
\frac{\partial P(x,v,t)}{\partial t} = {\bf{L}}P(x,v,t)
\end{equation}
where $\bf{L}$ stands for the so-called Fokker-Planck differential operator
\begin{equation}\label{eq:eq11}
\begin{array}{lll}
{\bf{L}}= {\bf{L}}(x,v)= &\frac{\partial }{\partial v}\left( {\frac{1}{\rho }\left( {\gamma (v)v + \nabla U(x)} \right) - \frac{1}{{{\rho ^2}}}G(v)G'(v)} \right) \\
&+ \frac{1}{{{\rho ^2}}}\frac{\partial }{{\partial v}}\left( {{{\left[ {G(v)} \right]}^2}\frac{\partial }{{\partial v}}} \right) - \frac{\partial }{{\partial x}}\left( v \right)
\end{array}
\end{equation}
The operator can be divided into two parts, a reversible $\bf{L}_{rev}$ and an irreversible $\bf{L}_{ir}$. The former is flux operator and the latter is collision operator, that is, 
\begin{equation}\label{eq:eq12}
\bf{L}={\bf{L}}_{rev}+{\bf{L}}_{ir}
\end{equation}
where $\bf{L}_{rev}$ and $\bf{L}_{ir}$ are given by
\begin{equation}\label{eq:eq13}
\left\{\begin{array}{lll}
{{\bf{L}}_{rev}} =  - v\frac{\partial }{{\partial x}} + \frac{{\omega _0^2}}{\rho }x\frac{\partial }{{\partial v}}\\
{{\bf{L}}_{ir}} = \frac{1}{\rho }\frac{\partial }{{\partial v}}\left( {\gamma (v)v + \frac{1}{\rho }{{\left[ {G(v)} \right]}^2}\frac{\partial }{{\partial v}}} \right)\\
\end{array}\right.
\end{equation}
Operator $\bf{L}(x,v)$ does not change with time $t$, so the solution of Eq.\ref{eq:eq10} is in exponential form. For the specific derivation, please refer to Ref.\cite{risken1984}. Therefore, the stationary solution of Eq.\ref{eq:eq10} can be obtained by multiplying the stationary solution of reversible and irreversible parts. The former can satisfy the Boltzmann distribution, it reads
\begin{equation}\label{eq:eq14}
P_{rev}^{st}(x,v) = {\rm{exp}}\left( {- \frac{{\omega _0^2{x^2}}}{{2\rho }} - \frac{{{v^2}}}{2}} \right)
\end{equation}

The collision operator $\bf{L}_{ir}$ is neither an anti-Hermitian operator nor a Hermitian operator. It satisfies
\begin{equation}\label{eq:eq15}
{\bf{L}_{ir}}P(x,v,t) = \frac{1}{\rho }\frac{\partial }{{\partial v}}\left( {\gamma (v)v + \frac{1}{\rho }{{\left[ {G(v)} \right]}^2}\frac{\partial }{{\partial v}}} \right)P(x,v,t) = 0
\end{equation}

Integrating Eq.\ref{eq:eq15} with respect to $v$, we yield 
\begin{equation}\label{eq:eq16}
\frac{\partial P(x,v,t)}{\partial v} =  - \frac{{\rho \gamma (v)v}}{{{{\left[ {G(v)} \right]}^2}}}P(x,v,t)
\end{equation}
This equation is the first-order variable coefficient differential one, so its formal solution can be obtained by integration, that is,
\begin{eqnarray}\label{eq:eq17}
\begin{array}{lll}
P_{ir}^{st}(x,v) = {\rm{exp}}\left( { - \rho {\gamma _0}\int {\frac{{({v^2} - \alpha )v}}{{2D{v^2} + 4\theta \sqrt {DQ} v + 2Q}}dv} } \right)\\
 ={\rm{exp}}\left( { - \rho {\gamma _0}\left( {\frac{{{v^2}}}{{4D}} - \frac{\theta }{D}\sqrt {\frac{Q}{D}} v + \frac{{(4{\theta ^2} - 1)Q - \alpha D}}{{4{D^2}}}ln\left( {{v^2} + 2\theta \sqrt {\frac{Q}{D}} v + \frac{Q}{D}} \right)} \right)} \right)\\
 \times {\rm{exp}}\left( { - \rho {\gamma _0}\theta \sqrt {\frac{Q}{D}} \left( {\frac{{(4{\theta ^2} - 1)Q - \alpha D}}{{2{D^2}}} + \frac{Q}{{{D^2}}}} \right)arctan\left( {\frac{{\sqrt {\frac{D}{Q}} v + \theta }}{{\sqrt {(1 - {\theta ^2})} }}} \right)} \right)
 \end{array}
\end{eqnarray}

It needs to point out that the strength of multiplicative noise $g(v)$ is selected as a linear function of velocity $v$ for simplicity and in a general case. So the stationary solution of Eq.\ref{eq:eq9} can take the following form 

\begin{equation}\label{eq:eq18}
\begin{array}{lll}
{P^{st}}(x,v) =& {\left( {{v^2} + 2\theta \sqrt {\frac{Q}{D}} v + \frac{Q}{D}} \right)^{ - \rho {\gamma _0}\left( {\frac{{(4{\theta ^2} - 1)Q - \alpha D}}{{4{D^2}}}} \right)}}\\
&\times {\rm{exp}}\left( { - \left( {\frac{{\omega _0^2{x^2}}}{{2\rho }} + \frac{{{v^2}}}{2}} \right) - \rho {\gamma _0}\left( {\frac{{{v^2}}}{{4D}} - \frac{\theta }{D}\sqrt {\frac{Q}{D}} v} \right)} \right)\\
 &\times {\rm{exp}}\left( { - \rho {\gamma _0}\theta \sqrt {\frac{Q}{D}} \left( {\frac{{(4{\theta ^2} - 1)Q - \alpha D}}{{2{D^2}}} + \frac{Q}{{{D^2}}}} \right){\rm arctan}\left( {\frac{{\sqrt {\frac{D}{Q}} v + \theta }}{{\sqrt {(1 - {\theta ^2})} }}} \right)} \right)
 \end{array}
\end{equation}

To get the general solution of Eq.\ref{eq:eq9}, Eq.\ref{eq:eq1} can be transformed into  
\begin{equation}\label{eq:eq19}
\frac{\rm d}{{\rm{d}t}}\left( {\begin{array}{*{20}{c}}
x\\
v
\end{array}} \right) =  - \left( {\begin{array}{*{20}{c}}
0&{ - 1}\\
{\frac{{\omega _0^2}}{\rho }}&{\frac{{\gamma (v)}}{\rho }}
\end{array}} \right)\left( {\begin{array}{*{20}{c}}
x\\
v
\end{array}} \right) + \left( {\begin{array}{*{20}{c}}
0\\
{\frac{1}{\rho}{\left[ {G(v)} \right]}\xi (t)}
\end{array}} \right)
\end{equation}
which describes a 2D Ornstein-Uhlenbeck process\cite{singh2018}. Thus, the drift matrix $\bf{\gamma}$ is given by
\begin{equation}\label{eq:eq20}
{\bf{\gamma }} = \left( {\begin{array}{*{20}{c}}
0&{ - 1}\\
{\frac{\omega _0^2}{\rho }}&{\frac{\gamma (v)}{\rho }}
\end{array}} \right),
\end{equation}
and the components of diffusion matrix $\bf{D}$ are
\begin{equation}\label{eq:eq21}
{D_{xx}} = {D_{xv}} = {D_{vx}} = 0,{D_{vv}} = \frac{{\left[ {G(v)} \right]}^2}{\rho ^2}
\end{equation}
Then the diffusion matrix $\bf{D}$ can be written as
\begin{equation}\label{eq:eq22}
{\bf{D}} = \left( {\begin{array}{*{20}{c}}
0&0\\
0&\frac{{\left[ {G(v)} \right]}^2}{\rho ^2}
\end{array}} \right)
\end{equation}
To simplify the matrix, $\bf{\gamma}$, which contains the drift and diffusion matrix, we introduce a complete biorthogonal set of $\bf{u}_i^{(\alpha)}$ and $\bf{v}_i^{(\alpha )}$($i = 1, 2$ and $\alpha=1, 2$),
\begin{equation}\label{eq:eq23}
\left\{\begin{array}{lll}
{{\bf{u}}^{(1)}} = \left( {\begin{array}{*{20}{c}}
{ - 1}\\
{{\lambda _1}}
\end{array}} \right)\\
{{\bf{u}}^{(2)}} = \left( {\begin{array}{*{20}{c}}
1\\
{ - {\lambda _2}}
\end{array}} \right)\\
{{\bf{v}}^{(1)}} = \frac{1}{{{\lambda _1} - {\lambda _2}}}\left( {{\lambda _2},1} \right)\\
{{\bf{v}}^{(2)}} = \frac{1}{{{\lambda _1} - {\lambda _2}}}\left( {{\lambda _1},1} \right)\\
\end{array}\right.
\end{equation}
with the orthonormality and completeness relations
\begin{equation}\label{eq:eq24}
\left\{\begin{array}{lll}
\sum\limits_\alpha  {{\bf{u}}_i^{(\alpha )}{\bf{v}}_j^{(\alpha )}}  = \delta _{ij}\\
{\bf{u}}_i^{(\alpha )}{\bf{v}}_i^{(\beta )} = \delta _{\alpha \beta }\\
\end{array}\right.
\end{equation}
where ${\lambda_\alpha}$ represents eigenvalue of the drift matrix, and it is satisfied with ${\lambda_1}\ne{\lambda_2}$. So the expansion of matrix $\bf{\gamma}$ in the above complete space reads
\begin{equation}\label{eq:eq25}
\left\{\begin{array}{lll}
{\gamma _{ij}}{\bf{u}}_j^{(\alpha )}={\lambda_\alpha }{\bf u}_i^{(\alpha )}\\
{\bf v}_i^{(\alpha )}{\gamma _{ij}} = {\lambda _\alpha}{\bf {v}}_j^{(\alpha )}\\
{\gamma_{ij}}=\sum\limits_\alpha{{\lambda_\alpha}{\bf {u}}_i^{(\alpha )}{\bf {v}}_j^{(\alpha )}} \\
\end{array}\right.
\end{equation}

By comparing Eq.\ref{eq:eq25} with the components of matrix $\bf{\gamma}$, we obtain
\begin{equation}\label{eq:eq26}
\left\{\begin{array}{lll}
{\lambda _1} + {\lambda _2} = \frac{{{\gamma _0}({v^2} - \alpha )}}{\rho }\\
{\lambda _1}{\lambda _2} = \frac{{\omega _0^2}}{\rho }\\
\end{array}\right.
\end{equation}
And then, we have
\begin{equation}\label{eq:eq27}
\left\{\begin{array}{lll}
{\lambda _1} = \frac{1}{2}\left( {\frac{{{\gamma _0}}}{\rho }({v^2} - \alpha ) - \sqrt {\frac{{\gamma _0^2}}{{{\rho ^2}}}{{({v^2} - \alpha )}^2} - \frac{{4\omega _0^2}}{\rho }} } \right)\\
{\lambda _2} = \frac{1}{2}\left( {\frac{{{\gamma _0}}}{\rho }({v^2} - \alpha ) + \sqrt {\frac{{\gamma _0^2}}{{{\rho ^2}}}{{({v^2} - \alpha )}^2} - \frac{{4\omega _0^2}}{\rho }} } \right)\\
\end{array}\right.
\end{equation}

The moments of variable can be calculated by introducing Green's function, $G_{ij}(t) = {\left[ {\rm {exp}(-\bf{\gamma}t)}\right]_{ij}}=\sum\limits_\alpha{{e^{-{\lambda_\alpha}t}}{\bf{u}}_i^{(\alpha )}{\bf{v}}_j^{(\alpha)}}$. The first-order moments can be written as
\begin{equation}\label{eq:eq28}
\left\{\begin{array}{lll}
x(t) = \left\langle x \right\rangle  = {\left[ {{\rm{exp}} ( - {\bf{\gamma }}t)} \right]_{11}}x + {\left[ {\exp ( - {\bf{\gamma }}t)} \right]_{12}}v\\
v(t) = \left\langle v \right\rangle  = {\left[ {{\rm{exp}} ( - {\bf{\gamma }}t)} \right]_{21}}x + {\left[ {\exp ( - {\bf{\gamma }}t)} \right]_{22}}v\\
\end{array}\right.
\end{equation}
and the $2$-order moments
\begin{equation}\label{eq:eq29}
\sigma _{ij} = 2\sum\limits_{\alpha ,\beta } {\frac{{1 -{\rm{exp}}\left( { - ({\lambda _\alpha } + {\lambda _\beta })t} \right)}}{{{\lambda _\alpha } + {\lambda _\beta }}}{D^{(\alpha ,\beta )}}u_i^{(\alpha )}u_j^{(\beta )}} 
\end{equation}
where $D^{(\alpha,\beta)}= v_k^{(\alpha )}{D_{kl}}v_l^{(\beta )}$. The elements of the $2$-order moment are given by
\begin{equation}\label{eq:eq30}
\left\{\begin{array}{lll}
\sigma _{11}(t) = {\left[ {\frac{{G(v)}}{{\rho ({\lambda _1} - {\lambda _2})}}} \right]^2}\left( {\frac{{{\lambda _1} + {\lambda _2}}}{{{\lambda _1}{\lambda _2}}} - \frac{{{\lambda _1}{e^{ - 2{\lambda _2}t}} + {\lambda _2}{e^{ - 2{\lambda _1}t}}}}{{{\lambda _1}{\lambda _2}}} + \frac{{4{\lambda _1}{\lambda _2}}}{{{\lambda _1} + {\lambda _2}}}\left( {{e^{ - ({\lambda _1} + {\lambda _2})t}} - 1} \right)} \right)\\
{\sigma _{12}}(t) = {\sigma _{21}}(t) = {\left[ {\frac{{G(v)}}{{\rho ({\lambda _1} - {\lambda _2})}}} \right]^2}{\left( {{e^{ - {\lambda _2}t}} - {e^{ - {\lambda _1}t}}} \right)^2}\\
{\sigma _{22}}(t) = {\left[ {\frac{{G(v)}}{{\rho ({\lambda _1} - {\lambda _2})}}} \right]^2}\left( {{\lambda _1}(1 - {e^{ - 2{\lambda _1}t}}) + {\lambda _2}(1 - {e^{ - 2{\lambda _2}t}}) + \frac{{4{\lambda _1}{\lambda _2}}}{{{\lambda _1} + {\lambda _2}}}\left( {{e^{ - ({\lambda _1} + {\lambda _2})t}} - 1} \right)} \right)\\
\end{array}\right.
\end{equation}

Since the second moment is positive, the inverse of matrix exists. The elements can be given by
\begin{equation}\label{eq:eq31}
\left\{\begin{array}{lll}
\left( {\bf\sigma ^{ - 1}} \right)_{11} = \frac{\sigma _{22}}{\rm{Det}\bf\sigma}\\
\left( {\bf\sigma ^{ - 1}} \right)_{12} = \left( \bf\sigma ^{ - 1} \right)_{21} =  - \frac{\sigma _{12}}{\rm{Det}\bf\sigma} \\
\left( {\bf\sigma ^{ - 1}} \right)_{22} = \frac{\sigma _{11}}{\rm{Det}\bf\sigma }\\
\end{array}\right.
\end{equation}
where ${\rm{Det}\bf\sigma}$ denotes the determinant of matrix, and can be written as ${\rm{Det}\bf\sigma}=\sigma_{11}\sigma_{22}-\sigma_{12}\sigma_{21}$. So the transition probability, the solution of Eq.\ref{eq:eq9} in the biorthogonal sets, can be obtained via the Fourier transform, that is, 
\begin{equation}\label{eq:eq32}
\begin{array}{lll}
&P(x,v,\tau |{x_0},{v_0},0)= \frac{1}{{2\pi \sqrt {{\rm{Det}}\bf\sigma } }}{\rm{exp}}\left( { - \frac{1}{2}{{\left[ {{\bf\sigma ^{ - 1}}(\tau )} \right]}_{11}}{{\left( {x - x(\tau )} \right)}^2}} \right.\\
&\left. { - {{\left[ {{\bf\sigma ^{ - 1}}(\tau )} \right]}_{12}}\left( {x - x(\tau )} \right)\left( {v - v(\tau )} \right) - \frac{1}{2}{{\left[ {{\bf\sigma ^{ - 1}}(\tau )} \right]}_{22}}{{\left( {v - v(\tau )} \right)}^2}} \right)
\end{array}
\end{equation}
Therefore, the probability density can be written as
\begin{equation}\label{eq:eq33}
P(x,v,\tau ;{x_0},{v_0},0) = P(\left.{x,v,\tau } \right|{x_0},{v_0},0) \times {P^{st}}(x,v)
\end{equation}

\section{Diffusion characteristic of particles}

Now let's apply the formalism to evaluate the probability $P(x,v,t;x_0,v_0,0)$ in the non-equilibrium system. It is worth noting that the spatial distribution of particles tends to be uniform under some certain parameters. However, the volecity distribution of particles transforms from a random distribution to a near-equilibrium distribution, and then transforms into an unimodal pattern. After a careful numerical investigation, the parameters are chosen as ${\omega _0} = 5.0 \times {10^{ - 4}}$, $\gamma _0 = 1.0$, $\lambda = 1.0 \times {10^{ - 3}}$, $\rho  = 0.5$, $\alpha = 0.5$, $D = 0.02$ and $Q = 0.1$. Based on the velocity distribution, we investigate the transport characteristics by calculating diffusion coefficient, viscosity coefficient and thermal conductivity coefficient.

\subsection{Distribution of velocity}

\begin{figure}[htbp]
\centering
\begin{minipage}[t]{0.49\linewidth}
    \centering
    \includegraphics[width=1.0\textwidth]{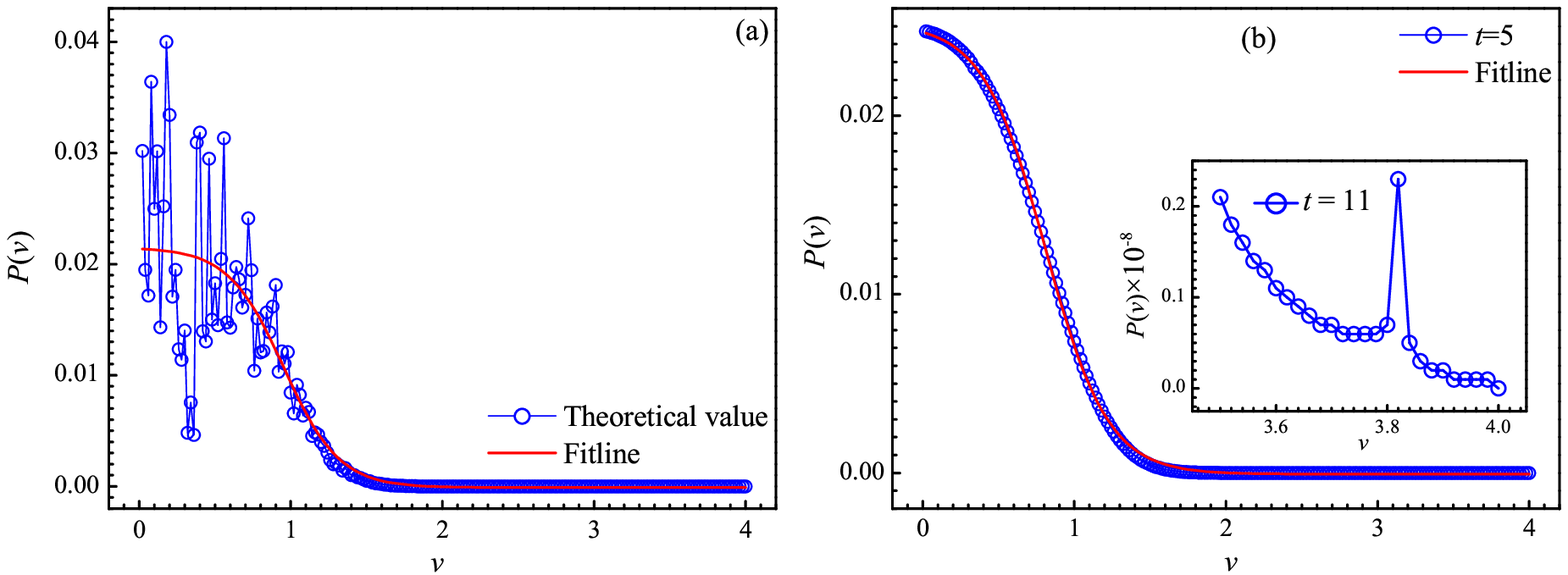}
\end{minipage}
\begin{minipage}[t]{0.49\linewidth}
    \centering
    \includegraphics[width=1.0\textwidth]{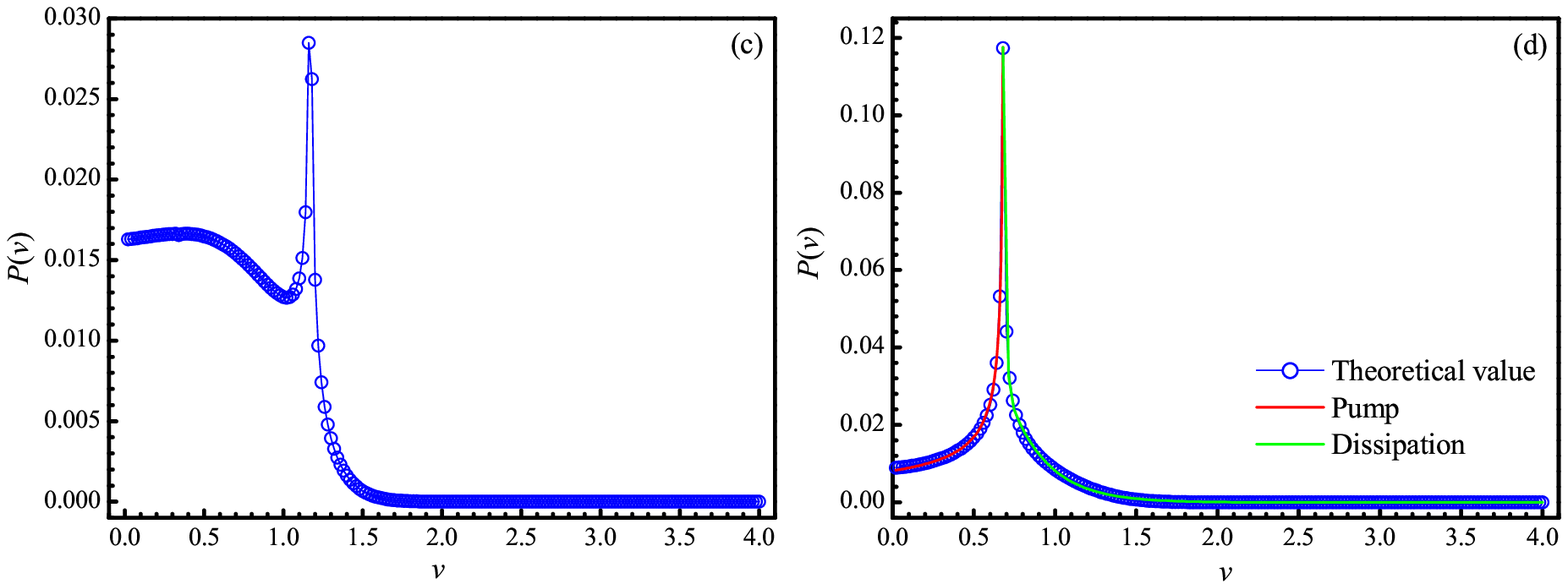}
\end{minipage}
\caption{The velocity distribution at different time step: (a) $t=1$, the red curve is the fitline; (b) $t=5$ and the inset(the partial enlargement) at $t=11$ ; (c) $t=150$; (d) $t=500$.}
\label{fig:fig1}
\end{figure}

Fig.~\ref{fig:fig1} presents the velocity distribution of particles. Fig.~\ref{fig:fig1}(a) shows that there is oscillation pattern at time $t=1$, which indicates that the velocity of particle is affected by random forces obviously at early stage. Fig.~\ref{fig:fig1}(b) presents that the velocity distribution, at $t = 5$, tends to a smooth near-equilibrium pattern. It is natural that the diffusion can eliminate the oscillation of distribution. It is worth noting that there is a spike occurring on the distribution curve when $t=11$, which indicates that the number of the particles, in the range covered by the spike, increases rapidly. Fig.~\ref{fig:fig1}(c) demonstrates that the spike is enlarged and moves to the left, and stops at $v_c=\sqrt{\alpha} = 0.7$(Fig.~\ref{fig:fig1}(d)). Obviously, the distribution curve is divided into two parts by $v_c$. The left one is the so-called pumping region as $v<v_c$, damping function $\gamma(v)<0$, while the right one is the dissipative region as $v>v_c$, $\gamma(v)>0$. As shown by Fig.~\ref{fig:fig1}(d), in the process of the spike moving in the dissipative region, some particles in the pumping region become faster due to the energy absorbing from the environment to form a single peak. This peak moves to the right and disappears at the boundary $v_c$ as the energy continuously injects, and finally forms the distribution dominated by a power law(the red fitting curve), $P(v)\propto (a-b\times v)^{-k_{p}}$. Meanwhile, the distribution in the dissipative region smoothly decays in the exponent manner(the green fitting curve), $P(v)\propto \exp{(-k_{e}v)}$. In one word, the increase of the particle energy in the pumping region and the decrease of that in the dissipative region jointly result in the sharp peak formed by the two smooth distribution curves. The detailed characteristics of the spike will be presented in the discussion on the energy distribution.  

\subsection{Distribution of kinetic energy}

Now let's discuss the distribution of energy to obtain more details of the system. The kinetic energy density of the system can be written as follows 
\begin{equation}\label{eq:eq34}
E = y(v) = \frac{1}{2}\rho v^2
\end{equation}
The probability presented by Eq. \ref{eq:eq33}, as $x$ and $v$ are the independent variables, can be transformed to the one that $x$ and $E$ are the independent variables, that is,
\begin{equation}\label{eq:eq34-1}
P(x,E,\tau)= \frac{\partial (x,v)}{\partial (x,E)}P(x,v,\tau;x_0,v_0,0)
\end{equation}
where the Jacobian matrix, $\frac{\partial (x,v)}{\partial (x,E)}$, in one dimension can be directly derived from Eq.\ref{eq:eq34},
\begin{equation}\label{eq3-34-2}
\frac{\partial (x,v)}{\partial (x,E)}=\frac{{\rm{d}}v}{{\rm{d}}y(v)}=\frac{1}{\sqrt{2\rho E}}
\end{equation}
So the probability density of kinetic energy can be formulated as
\begin{equation}\label{eq:eq35}
P(x,E,\tau) = {\sum\limits_n {P(x,y_n^{ - 1}(E),\tau )\left. {{{\left| \frac{dy(v)}{dv} \right|}^{ - 1}}} \right|} _{v = y_n^{ - 1}(E)}}
\end{equation}
where the inverse functions are $y_{1,2}^{-1}(E)= \pm \sqrt {\frac{{2E}}{\rho }}$, then Eq.\ref{eq:eq35} becomes
\begin{equation}\label{eq:eq36}
\begin{array}{lll}
P(x,E,\tau ) =& \sqrt {2\rho E} \left( {P(x, - \sqrt {\frac{2E}{\rho }} ,\tau ;{x_0}, - \sqrt {\frac{2{E_0}}{\rho }} ,0)} \right.\\
 &\left.+ P(x,\sqrt {\frac{2E}{\rho }} ,\tau ;{x_0},\sqrt {\frac{2{E_0}}{\rho }},0)\right)
 \end{array}
\end{equation}

\begin{figure}[htbp!]
\centering
\begin{minipage}[t]{0.49\linewidth}
    \centering
    \includegraphics[width=1.0\textwidth]{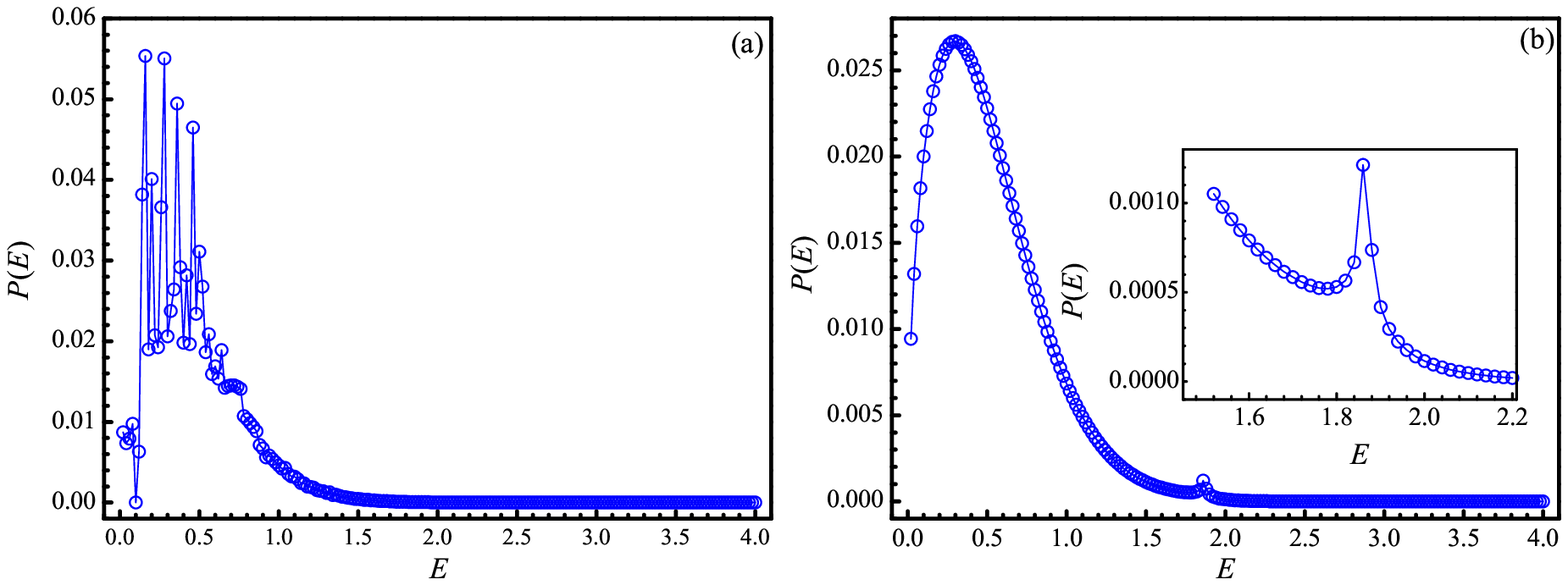}
\end{minipage}
\begin{minipage}[t]{0.49\linewidth}
    \centering
    \includegraphics[width=1.0\textwidth]{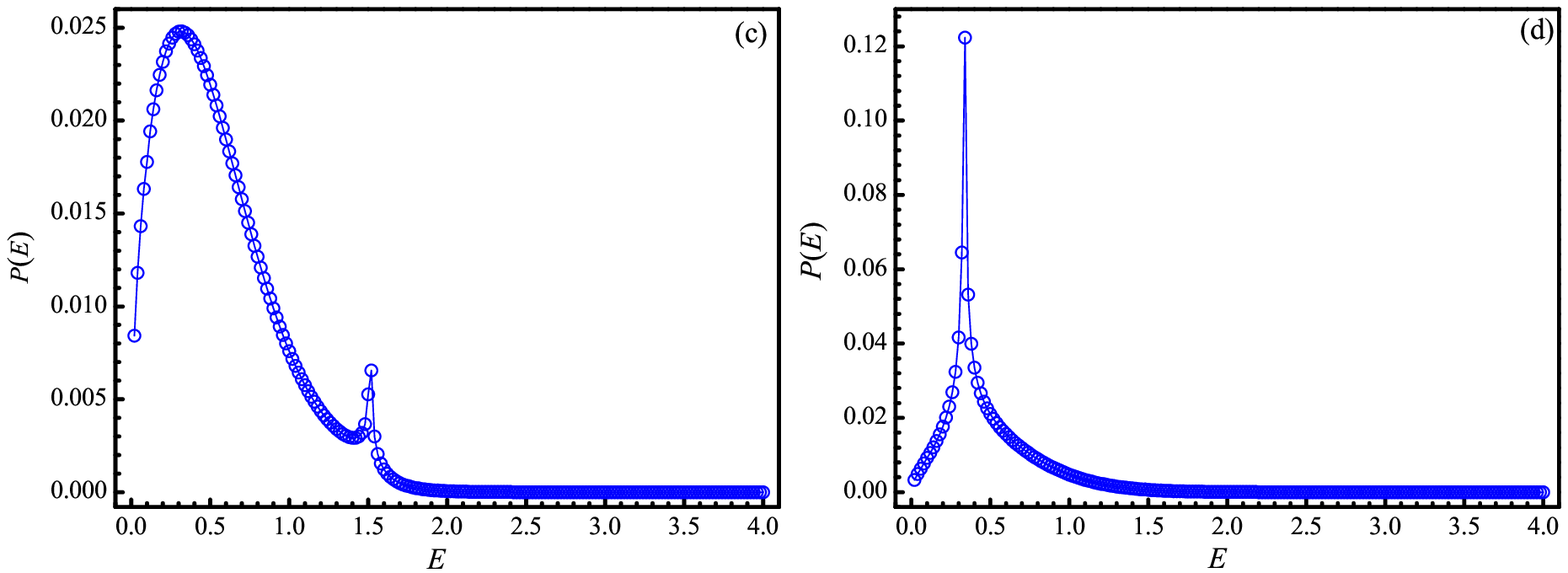}
\end{minipage}
\caption{The distribution of energy at different time: (a) $t=1$; (b) $t=30$, the inset is the partial magnification; (c) $t=150$; (d) $t=500$.}
\label{fig:fig2}
\end{figure}

Fig.~\ref{fig:fig2} re-shows the statistical characteristics of the system by the energy distribution instead of the velocity distribution shown by Fig.~\ref{fig:fig1}. Please note that Fig.~\ref{fig:fig2}(b) presents the non-equilibrium spike when $t=30$ at which it has developed for $19$ time units after the occurrence($t=11$). In fact, the spike moving left in the dissipative region can be attributed to the energy dissipation caused by the collision between particles, which leads to the number of particles with the peak energy decreases, but that of the particle with the energy being slightly lower and near to the peak energy increases. The kinetic energy space is also divided into the pumping region $E < E_c$ and the dissipative region ${E} >E_c$, where $E_c=0.49$ that corresponds to $v_c$. The two smooth curves form a sharp peak of energy distribution, for the same reason discussed on the peak profile in Fig.~\ref{fig:fig1}(d). Now we can interpret the oscillation in Fig.~\ref{fig:fig1}(a) precisely. We find that the oscillation mainly exists in the pumping region, and the energy absorbing of the particle enlarges the fluctuations in the distribution of energy. In the process from the situation shown by Fig.~\ref{fig:fig2}(b) to that shown by Fig.~\ref{fig:fig2}(d), a remarkable profile of the spike shown by the inset of Fig.~\ref{fig:fig2}(b)(the enlargement of the spike) and Fig.~\ref{fig:fig2}(c) is asymmetrical. Obviously, the gradient of the pre-peak is greater than the absolute value of the gradient of the post-peak. So the pre-peak takes on the form of step function, and the post-peak decays monotonously. The monotonously decreasing post-peak means that while the energy dissipates, the spike is being nonlinearly enlarged to sustain. The first descending and then ascending pre-peak is produced by the energy transport via the collision between particles, which leads to the rapid increase of number of the particles with energy near to the peak point and the corresponding decrease of the number of the particles with lower energy. So the pre-peak ascends steeply. Finally, the increase of the particle energy in the pumping region and the decrease of that in the dissipative region jointly result in the sharp peak formed by the two smooth distribution curves. This process is similar to the shock process formed by the diffusion such as a detonation or deflagration\cite{gan2015,Khokhlov1997}. 

\begin{figure}[htbp!]
   \centering
   \includegraphics[width=0.4\textwidth]{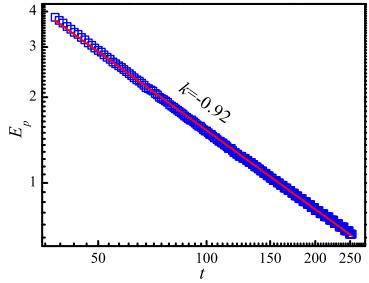}
   \caption{Power scaling relation between the position of the peak value and time. }
   \label{fig:fig3}
\end{figure}

Fig.~\ref{fig:fig3} presents the dependence of the position of the peak value, $E_{\rm{p}}$, of the spike on time $t$, which shows better agreement with power law, $E_{\rm{p}}\propto t^{-k}$, where $k=-0.92$. This scaling law reveals two basic facts: one is that the relaxation time of collision, ${\tau _0} = \frac{l_0}{v_{p}}$, between particles becomes long with fixed free path $l_{0}$, due to the energy dissipating and velocity deceasing; the other is that the energy transport slows down with time in the power law manner, and the kinetic energy converts into the intrinsic energy of the system.

\subsection{Characteristics of the correlations}

Although we have tried to describe the non-equilibrium spike above, there are still some problems to interpret. What causes the non-equilibrium spike? Why does it occur in the high-energy region?

To answer the questions, we discuss the correlations among the relevant variables. It's well known that the transport is driven by the gradient force for an ideal gas system, while that is related to the interactions among particles in addition to the gradient in a real system. To look for the origin what causes the non-equilibrium spike, it is necessary to calculate the moments of variables and the correlations among variables. As we all know, the force existing in the system can be classified into two type, macro-force and micro-force. Here the former is the gradient force, and the latter is the damping force and interaction force caused by collision among particles. By calculating these forces, we find that they all decays with time and can not directly induce non-equilibrium behavior. So we calculate the correlation strength between them and the space, as well as the 3-order moment of velocity. They are defined as 
\begin{equation}\label{eq:eq37}
\left\{\begin{array}{lll}
\Delta ^G = \left\langle {\nabla U(x(t)) \cdot x(t - \tau )} \right\rangle \\
\Delta ^D = \left\langle {-{\gamma _0}(v{{(t)}^2} - \alpha )v(t) \cdot x(t - \tau )} \right\rangle \\
\Delta ^R = \left\langle  {G(v(t))} \xi (t) \cdot x(t - \tau ) \right\rangle \\
\Delta ^{3,v} = \left\langle (v-\left\langle v\right\rangle)^3 \right\rangle
\end{array}\right.
\end{equation}
where $\Delta^G$, $\Delta^D$ and $\Delta^R$ denote the correlation strength between the field force, damping force, random force and the space, respectively. $\Delta^{3,v}$ denotes the $3$-order moment of velocity.

\begin{figure}[htbp!]
   \centering
\begin{minipage}[t]{0.49\linewidth}
    \centering
    \includegraphics[width=1.0\textwidth]{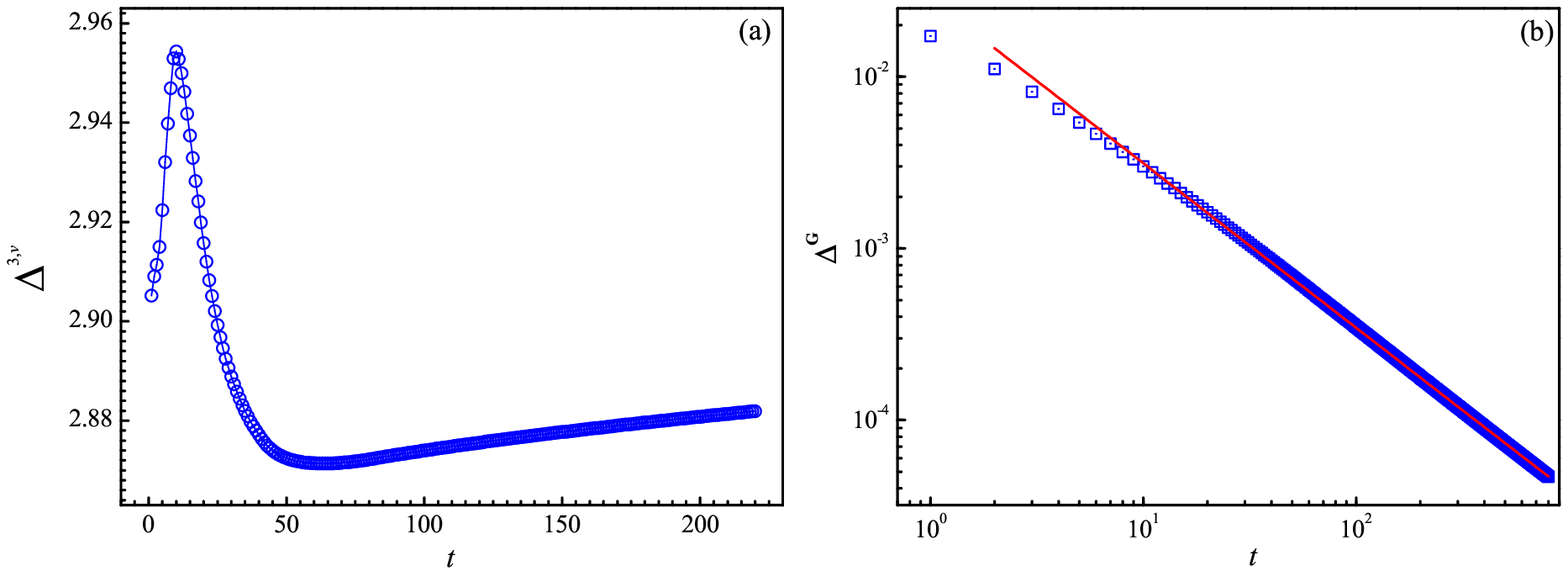}
\end{minipage}
\begin{minipage}[t]{0.49\linewidth}
    \centering
    \includegraphics[width=1.0\textwidth]{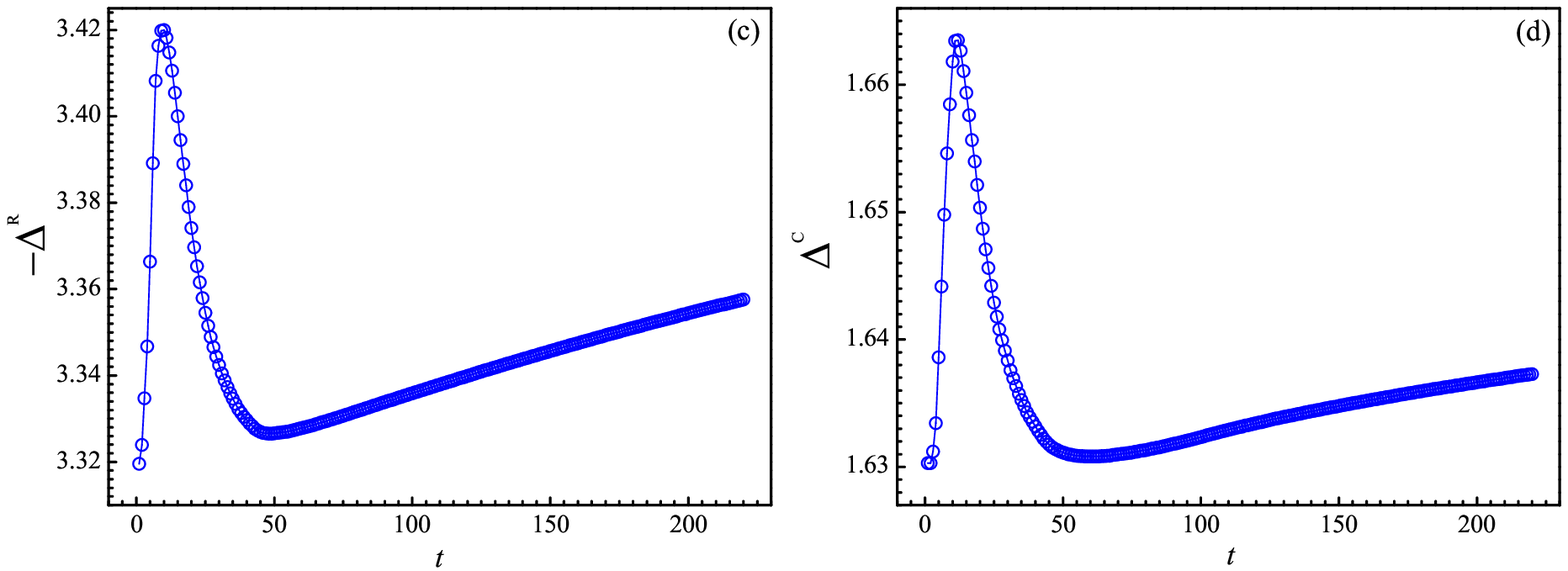}
\end{minipage}
   \caption{The variations of moment and correlations with time: (a) the $3$-order moment shows the non-equilibrium effect at time $t=11$; (b) the correlation between macro-force and space monotonously decays with time; (c) the minus correlation between damping force and the space shows a pulse at time $t=11$; (d) the correlation between noises and space induces the non-equilibrium effect.}
   \label{fig:fig4}
\end{figure}

Fig.~\ref{fig:fig4} presents the variation of the $3$-order moment of velocity, as well as those of the correlations between the forces and the space with time. It is known that the $3$-order moment of velocity, named as skewness, is the measurement of the deviation degree of the distribution from the normal distribution. Fig.~\ref{fig:fig4}(a) shows that after a monotonous increase the skewness, $\Delta^{3,v}$, reaches the maximum at time $t=11$ when the non-equilibrium spike initially appears. One may conclude that the spike interrupts the increase of skewness. In other words, it is the sign for the spike that the skewness stops the monotonous variation. What causes the non-equilibrium spike? It is certainly not $\Delta^{G}$ and $\Delta^{D}$ due to the fact that the former decays with time in a power law(Fig.~\ref{fig:fig4}(b)) and the latter varies following a valley(Fig.~\ref{fig:fig4}(c)). It is only $\Delta^{R}$ because it varies in the way similar to that of $\Delta^{3,v}$, that is, the value also reaches the maximum at the same time, $t=11$(Fig.~\ref{fig:fig4}(d)). Furthermore, it seems that there is a threshold value of correlation $\Delta^{R}$, the maximum, and as the correlation is strengthened to the threshold the non-equilibrium may be triggered. In fact, from Eq.\ref{eq:eq8} and $g(v)=v$, we know the noise is positively correlated with the velocity. Therefore, the noise is introduced deeply into the system by the correlation between the noise and the space, and enlarged nonlinearly. As discussed above, the positive skewness reflects the three aspects of the non-equilibrium effect. The first one is that the effect originates from the correlation between the noise and the space; the second one is the dissipation of energy; and the third one is the ability of the system to enlarge the random fluctuation, which can be inferred from the fact that the spike is not worn out, but it is preserved until it meets the pumping region and forms the sharp peak distribution. It is natural that the correlation of noise with the space at which there is larger velocity/energy could be enlarged more easily. Therefore, the non-equilibrium spike will emerge in higher velocity/energy region.

\subsection{Characteristics of the transport}

To reveal the detailed patterns of particle diffusion, we investigate the statistical features via calculating diffusion coefficient, viscosity coefficient and thermal conductivity coefficient.

The mean square displacement is given by   
\begin{equation}\label{eq:eq38}
\left\{\begin{array}{lll}
\left\langle x \right\rangle  = \int\limits_{x_0}^{x_{\rm{max}}} {\int\limits_{v_{\rm{min}}}^{v_{\rm{max}}}{xP(x,v,\tau ;{x_0},{v_0},0)dxdv}} \\
\left\langle {{{\left( {x - \left\langle x \right\rangle } \right)}^2}} \right\rangle =\int\limits_{x_0}^{x_{\rm{max}}} {\int\limits_{v_{\rm{min}}}^{v_{\rm{max}}}{x^2P(x,v,\tau ;{x_0},{v_0},0)dxdv}}- {\left\langle x \right\rangle ^2}\\
\end{array}\right.
\end{equation}

Based on the Fick's second law, the effective diffusion coefficient is given by\cite{megen1986,megan1987,pusey1982,douglas1997}
\begin{equation}\label{eq:eq39}
D_{\rm{eff}} = \frac{\left\langle {\left(x - \left\langle x \right\rangle \right)}^2 \right\rangle }{2dt}
\end{equation}
where $d$ denotes the spatial dimension, which is chosen as $d=1$ here. 

\begin{figure}[htbp!]
   \centering
   \includegraphics[width=0.8\textwidth]{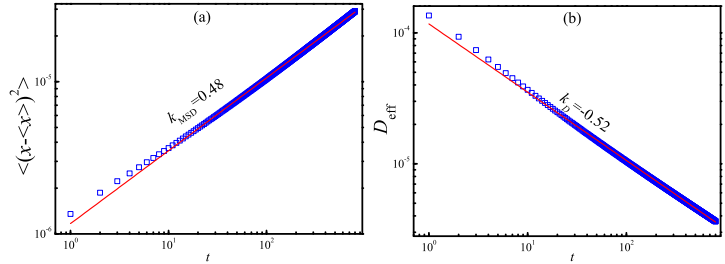}
   \caption{The scaling law dominating the variations of the mean square displacement and effective diffusion coefficient with time on log-log coordinate plane: (a) the scaling law for mean square displacement; (b) the scaling law for effective diffusion coefficient.}
   \label{fig:fig5}
\end{figure}

Fig.~\ref{fig:fig5} shows the variations of the mean square displacement and effective diffusion coefficient with time. The red curve in Fig.~\ref{fig:fig5}(a) demonstrates the fitting by power function, $\left\langle {\left(x - \left\langle x \right\rangle \right)}^2 \right\rangle \propto t^{k_{\rm{MSD}}} (k_{\rm{MSD}}=0.48<1)$, which implies that the particles undergo sub-diffusion as a result of the viscous resistance. Fig.~\ref{fig:fig5}(b) presents the variation of the effective diffusion coefficient with time, which can also be fitted by a power law, $D_{\rm{eff}}\propto t^{-k_{\rm{D}}} (k_{\rm{D}}=-0.52)$. This scaling relation further proves the sub-diffusion.

Next, we calculate the viscosity coefficient and the thermal conductivity coefficient. The former is defined by the correlation function between momentum and the space,
\begin{equation}\label{eq:eq41}
\eta  = \frac{1}{2dLt}\left\langle {\left( {\rho vx - \left\langle {\rho vx} \right\rangle } \right)}^2 \right\rangle
\end{equation}
where $L$ is the spatial scale. Similarly, the latter is defined by the correlation between the energy and the space,
\begin{equation}\label{eq:eq41-2}
\kappa = \frac{1}{2d{T^2}Lt}\left\langle {\left( {xe - \left\langle {xe} \right\rangle } \right)}^2 \right\rangle 
\end{equation}
where $T$ is the thermodynamics temperature, $e$ the kinetic energy. 

\begin{figure}[htbp!]
   \centering
   \includegraphics[width=0.8\textwidth]{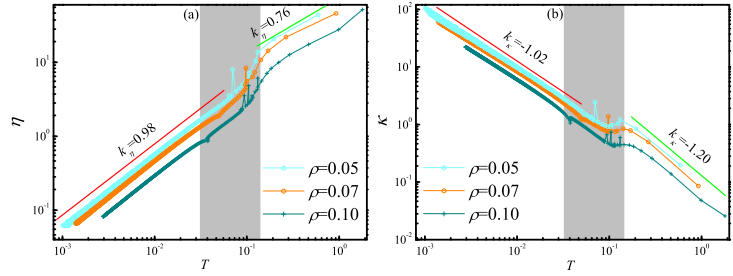}
   \caption{The dependence of the transport parameters on temperature: (a) the double-scaling relation between the viscosity coefficient and temperature, the power scaling exponents are $k_{\rm{\eta}}=0.99$ and $k_{\rm{\eta}}=0.75$, respectively; (b) the double-scaling relation between the thermal conductivity coefficient and temperature, the power scaling exponents are $k_{\rm{\kappa}}=-1.02$ and $k_{\rm{\kappa}}=-1.20$, respectively.}
   \label{fig:fig6}
\end{figure}

In calculating, the value of the spatial scale is chosen as $L=2.0$. Fig.~\ref{fig:fig6} shows the double-scale power relation between viscosity coefficient and temperature, and that between thermal conductivity coefficient and temperature. It is easy to understand this double-scaling phenomenon, that is, the scaling law established initially in the near-equilibrium stage, described by the Boltzmann distribution, is destroyed by the non-equilibrium spike, and then a new scaling law re-establishes when the transport induced by the correlation between noise and space is fulfilled and the final sharp peak distribution is formed. So there is a transition zone between the two scaling shown in Fig.~\ref{fig:fig6}. We note that the transition zones at different densities show that the bigger the density, the bigger the transition zone is, and the more difficult the new scaling is re-established due to the fact that the more serious damage to the previous scaling, the more difficult influence can be eliminated.

\section{The propagation characteristics in a social system}

Human behavior, especially the phenomena of public opinion propagation connected with human activities in society, are in far-from equilibrium state, and such a system are of typical open one\cite{bartolozzi2005,Jiang2008}. We may be permitted to take the agents in social system as the Brownian particles based on the similarities between the two, stochastic motion, interaction mode, and so on\cite{Schweitzer2000}. So the results obtained in the aforementioned physical system can be cast light upon the social systems to understand thier behaviors deeply. 

\subsection{Applying the random dynamics model to a social system}

Most of all that each of the variables and parameters describing the particle system should be corresponded to the social system based on the presupposition that the opinion particle obeys stochastic dynamics described by Eq.\ref{eq:eq4}. Being consistent with the particles, the opinion particles should also carry some physical quantities such as energy and momentum. Here the opinion particle stands for any one of the agents of social system in an opinion exchange process or a message propagation process. From the perspective of physics, the fundamental variables to describe the state of an opinion particle are the displacement and velocity. Inspired by the conceptions of social space and opinion space respectively suggested in Ref.\cite{Lewenstein1991} and Ref.\cite{Pineda2009,fortunato2005}, we propose an information space in which the position of an opinion particle relative to the origin is used to define the generalized displacement. That is to say the generalized displacement actually is the information quantity carried by the opinion particle.  The reason for this is that the information captured by an individual with high cognitive ability from an event is generally more than that one with a low cognitive ability. The velocity, the changing rate of the generalized displacement/information with respect to time may be looked upon as the sensitivity of the opinion particle to an event. Besides, the other variables in Eq.\ref{eq:eq4} are  endowed with some special meaning. The positive damping implies the decrease of information as the energy of opinion particle is dissipated and the event gradually fades from the public eye, while the negative damping means the increase of information as energy is injected into the system to strengthen the message propagation. The social noise is actually the random influence exerted on the opinion particle by the internal and/or external environment via the correlation of it with the generalized displacement. Here, the correlation refers to the randomly occurred event has been concerned. 

\subsection{The characteristic of opinion propagation}

In the physical system, we find that the correlation between the noise and the space induces the non-equilibrium spike in the sub-diffusion process. The non-equilibrium spike moves in the direction of velocity decreasing, and ends up at the border between the pumping and dissipative regions. Corresponding to Fig.~\ref{fig:fig1}, the opinion propagation process with the shift of non-equilibrium spike can be understood as follows. As shown in Fig.~\ref{fig:fig1}(a), in the preliminary stage information exchanges among the individuals will cause big fluctuations and therefore a confuse due to the fact that most of individuals may know the news but don't know the truth, which indicates that an exchange of information or a message-sharing might lead to the turbulent distribution. After all, the system tends to be in a stable state, the near-equilibrium state shown by Fig.~\ref{fig:fig1}(b), after the full information exchange among the individuals. Naturally, this "quiescent" equilibrium state will be broken by a sudden attention, comment, and/or public sentiment which may be taken as a kind of social noise and play a part via its correlation with information(the correlation here means in essence that the noise permeates through the system, and become part of the propagating information). So such information will propagate in the "susceptible population" those are sensitive to the noise and with great velocity, which causes the spike(presenting in Fig.~\ref{fig:fig1}(b)) moving in the direction of the sensitivity decreasing(Fig.~\ref{fig:fig1}(c)). This spike likes some event spreading on the web, for instance, the event causes a peak in space of attention number-time\cite{wang2017,liu2017} and the intermittent characteristics of the access to web site\cite{goncalves2008}. Generally speaking, the opinion particles in the pumping region are the ones insensitive to the news or message. The pumping mechanism could be regarded as the external influence or guiding learning help provided for the individuals of poor-learning ability or insensitive to things to efficiently improve their ability. Please note, in this process, the curve of the velocity distribution transformed from the convex one into the concave one and finally forms the rapid growth power law. From the discussion above, we may draw a conclusion that some measures such as government supporting, guiding learning and so on should be taken to rapidly improve ability of the individuals who are of poor-learning ability or insensitivity to new thing, while although the noise, in social system, such as some public opinions caused by lies and slanders will be dissipated and eliminated, they may be nonlinearly enlarged in the propagation process and contribute partly to the sharp peak of the distribution, and therefore their great harmfulness to society should be a matter of vigilance.

\begin{figure}[htbp!]
   \centering
   \includegraphics[width=0.4\textwidth]{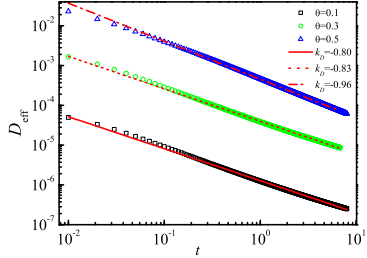}
   \caption{The dependence of the diffusion coefficient on time at different $\theta$ obey power law.}
   \label{fig:fig7}
\end{figure}

The sub-diffusion of the opinion particles in the system implies that the propagation of information is impeded. We also find that there is a negative correlation between the scaling exponent $k_{\rm{D}}$ in the relation of the diffusion coefficient to time and the intensity factor of noise-correlation $\theta$. As is shown by Fig.~\ref{fig:fig7} that the scaling exponent $k_{\rm{D}}$ equals to $-0.80$, $-0.83$ and $-0.96$ in the case of $\theta=0.1,0.3$ and $0.5$, respectively. When $\theta>0.5$, the power scaling exponent $k_{\rm{D}}<-0.96$, and is close to $-1$, which implies that the scaling exponent of mean square displacement $k_{\rm{MSD}}$ is close to $0$, so the mean square displacement keeps changeless. It is known that the fixed  mean square displacement corresponds to a uniform distribution in the space, which is essentially the limit case of the sub-diffusion---the distribution becomes uniform finally. Of course, this is a natural result due to the dissipation mechanism that the spike induced by a social noise will eventually die away. The empirical observations mentioned in Ref.\cite{wang2017,liu2017} are the typical examples. 

\section{Conclusions and Discussions}

In this paper, we investigate the transport behaviors in an open system composed of classical particles. A Langevin equation is suggested to express the stochastic dynamic process in which the additive and multiplicative noises as well as the positive and negative damping effect are synchronously stressed. The equation is transformed into the Fokker-Planck equation by means of the Kramers-Moyal expansion and the Green function. The solution brings to light the details of the transport behavior, which is also mapped to the social system formed by opinion particles to describe the information propagation process. 

In particle system, the early distribution of the velocity/energy is greatly subjected to random forces, especially in the lower velocity/energy region, so the distribution curve oscillates drastically. The oscillation may disappear quickly, and sets up the near-equilibrium state described by the Boltzmann distribution. The near-equilibrium distribution is destroyed by a non-equilibrium spike at the location where velocity/energy is high in the dissipative region due to the correlation of the noise and the generalized displacement. The spike moves in the direction of velocity/energy decreasing due to the energy dissipation mechanism. The energy is compensated, to a certain extent, by the spike itself via being nonlinearly enlarged. The continuation of this trend leads to the exponentially decaying distribution in the dissipative region, meanwhile the energy injection under the pumping mechanism results in the power law distribution in the pumping region, and therefore the abutment of the two regions forms the sharp peak of the distribution---one new near-equilibrium state. The distribution of energy can provide more detailed description to the diffusion behavior. The power laws dominating the dependence of the mean square displacement on time with scaling exponent $0.48<1$, and that of the diffusion coefficient on time with scaling exponent $-0.52<0$ jointly determine that the diffusion is substantially of a sub-diffusion. The power function relating the viscosity coefficient to temperature with positive scaling exponent and that relating the thermal conductivity coefficient to temperature with negative scaling exponent prove that this transport process is similar to that undergone in the macromolecule or polymer such as clathrate hydrate\cite{krivchikov2006,krivchikov2005,andersson1996}, single crystal\cite{yang2010,alshaikhi2010}, protein particles\cite{savin2014}.

Based on the similarity between the diffusion of particles in physical system and the propagation of message in social system, the description of the physical system is mapped to the physical description to the social system. Most important of all, the generalized displacement in a social system is taken as the quantity of information carried by the opinion particle, and the velocity is regarded as the sensitivity of the opinion particle to an event. The non-equilibrium spike arises on the background of the near-equilibrium distribution can be regarded as the the message randomly introduced into the system and the individuals of higher sensitivity to the noise are generally the first ones that are influenced. The message propagates and gradually affects the individuals of lower sensitivity to the message, so the dissipation and the nonlinear magnification mechanisms allow the spike moves in the direction of energy decreasing. Meanwhile, the pumping mechanism impels the individuals of previously with lower cognitive ability improve their sensitivity to things. These findings may remind us that the assistance from social institutions to the individuals of lower cognitive ability is important and will take effect rapidly, while the tattle and prate as the noise introduced into the social system may be nonlinearly enlarged.

\section{Acknowledgments}
This study is supported by National Natural Science Foundation of China(Grant No. 11665018) and The Major Innovation Projects for Building First class Universities in China's Western Region (Grant No. ZKZD2017006).

\end{document}